\PassOptionsToPackage{svgnames}{xcolor}
\documentclass[preprint, journal]{vgtc}                     

\onlineid{0}

\vgtccategory{Research}
\vgtcpapertype{application/design study}

\title{PUREsuggest: Citation-based Literature Search and Visual Exploration with Keyword-controlled Rankings}

\author{\authororcid{Fabian Beck}{0000-0003-4042-3043}}

\authorfooter{
    \item Fabian Beck is with University of Bamberg. \\
    E-mail: fabian.beck@uni-bamberg.de

}

\abstract{%
Citations allow quickly identifying related research. If multiple publications are selected as seeds, specific suggestions for related literature can be made based on the number of incoming and outgoing citation links to this selection. Interactively adding recommended publications to the selection refines the next suggestion and incrementally builds a relevant collection of publications. Following this approach, the paper presents a search and foraging approach, PUREsuggest, which combines citation-based suggestions with augmented visualizations of the citation network. The focus and novelty of the approach is, first, the transparency of how the rankings are explained visually and, second, that the process can be steered through user-defined keywords, which reflect topics of interests. The system can be used to build new literature collections, to update and assess existing ones, as well as to use the collected literature for identifying relevant experts in the field. We evaluated the recommendation approach through simulated sessions and performed a user study investigating search strategies and usage patterns supported by the interface.  %
} 


\teaser{
  \centering
  \includegraphics[width=\linewidth, trim=0.3in 0.3in 0.3in 0.4in, clip=true]{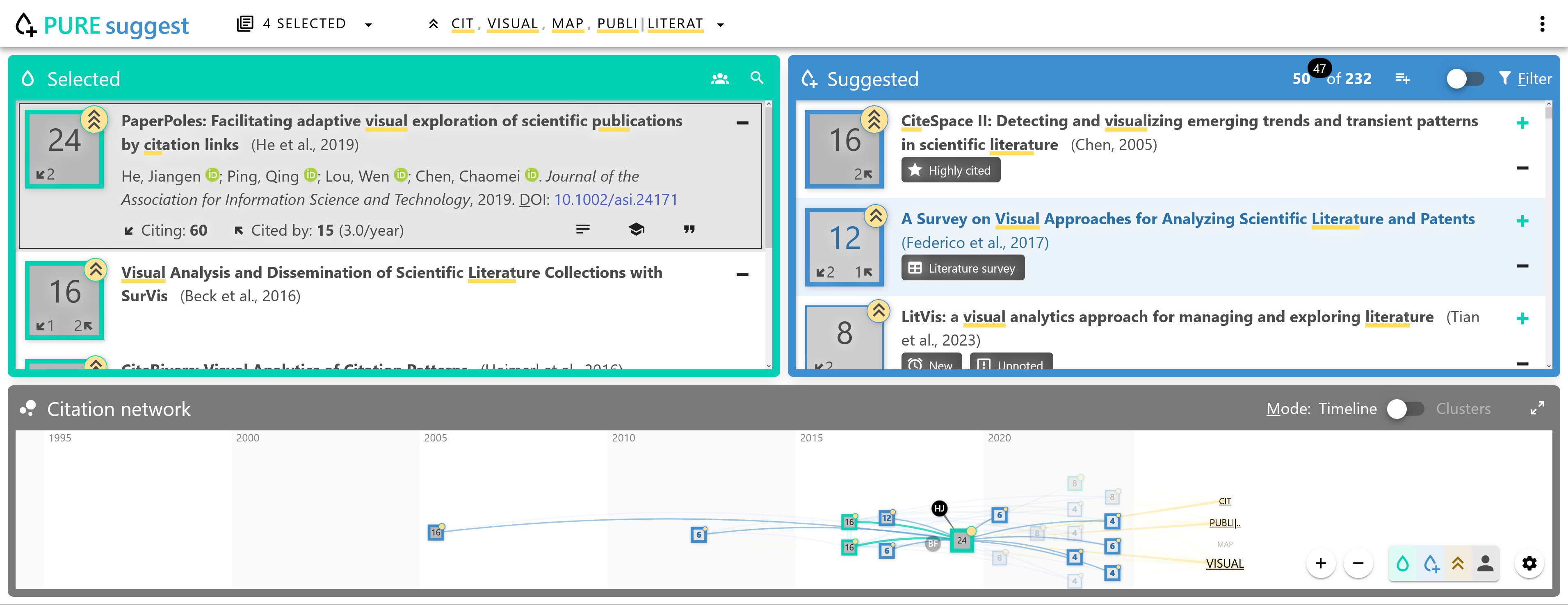}
  \caption{The interface of \toolname consists of three main components: \cmarker{color_selected} the list of currently selected publications, \cmarker{color_suggested}~the list of suggestions made on the basis of the selection, and \cmarker{color_network} a visualization of the respective citation network (here, shown as a \textit{timeline}); \cmarker{color_keyword}~user-defined boost keywords that amplify topics of interest are entered at the top and highlighted across the interface.
  }
  \label{fig:teaser}
}

\graphicspath{{figs/}{figures/}{pictures/}{images/}{./}} 

\usepackage{mathptmx}                  

\usepackage{amsmath}                        
\usepackage{amsfonts}                       
\usepackage{amssymb} 

\usepackage[usenames,dvipsnames]{color}

\newcommand{\tablefontsize}{\fontsize{6.4}{8}\selectfont}

\usepackage{booktabs}
\usepackage{tabularx}
\usepackage{array}
\usepackage{multirow}
\usepackage{colortbl}
\renewcommand{\arraystretch}{1.3}

\newcolumntype{L}{>{\raggedright\arraybackslash}X}
\newcolumntype{C}{>{\centering\arraybackslash}X}
\newcolumntype{R}{>{\raggedleft\arraybackslash}X}

\usepackage{pgf}
\newcommand{\ccell}[1]{%
    \pgfmathsetmacro{\grayvalue}{#1*4} 
    \xdef\tempcolor{black!\grayvalue}   
    \colorbox{\tempcolor}{$#1$}            
}

\usepackage{sparklines}
\newcommand{\factor}{0.025} 
\newcommand{\evalsparkline}[3]{
    \begin{sparkline}{3}
        \definecolor{sparkspikecolor}{named}{red}
        \pgfmathsetmacro{\valone}{#1*\factor}
        \sparkspike .1 {\valone}
        \definecolor{sparkspikecolor}{named}{gray}
        \pgfmathsetmacro{\valtwo}{#2*\factor}
        \sparkspike .48 {\valtwo}
        \definecolor{sparkspikecolor}{named}{blue}
        \pgfmathsetmacro{\valthree}{#3*\factor}
        \sparkspike .86 {\valthree}
    \end{sparkline}
}
\newcommand{\factorus}{0.19} 
\newcommand{\evalsparklineus}[5]{
    \begin{sparkline}{3}
        \definecolor{sparkspikecolor}{named}{red}
        \pgfmathsetmacro{\valone}{#1*\factorus}
        \sparkspike .1 {\valone}
        \definecolor{sparkspikecolor}{HTML}{F49EC4}
        \pgfmathsetmacro{\valtwo}{#2*\factorus}
        \sparkspike .46 {\valtwo}
        \definecolor{sparkspikecolor}{named}{gray}
        \pgfmathsetmacro{\valthree}{#3*\factorus}
        \sparkspike .82 {\valthree}
        \definecolor{sparkspikecolor}{HTML}{8787FF}
        \pgfmathsetmacro{\valfour}{#4*\factorus}
        \sparkspike 1.18 {\valfour}
        \definecolor{sparkspikecolor}{named}{blue}
        \pgfmathsetmacro{\valfive}{#5*\factorus}
        \sparkspike 1.54 {\valfive}
    \end{sparkline}
}

\usepackage{pifont}

\newcommand{\spartly}[0]{$\circ$\xspace}
\newcommand{\syes}[0]{$\bullet$\xspace}

\newcommand{\inlinegraphic}[1]{%
	\protect\raisebox{-1pt}{%
		\protect\includegraphics[height=0.28cm]{inlinegraphics/#1}%
	}\,%
}

\newcommand{\linegraphicdouble}[1]{%
    \vspace{0.1cm} \\ \vspace{0.1cm} \centerline{
        \includegraphics[height=0.6cm]{figures/#1}%
    } \\
}

\newcommand{\fgraphic}[2]{%
    \setlength{\fboxsep}{0pt}
    \fbox{\includegraphics[#1]{figures/#2}}
}

\usepackage{xhfill}

\newcommand{\cmarker}[1]{\raisebox{2.5pt}{\fcolorbox{white}{#1}{\rule{0pt}{0pt}}}}

\usepackage{xspace} 
\newcommand{\toolname}{\textit{PUREsuggest}\xspace}

\newcommand{\vtransparancy}{{\color{color_value}\small\sffamily\textbf{T\&E}}}
\newcommand{\vcontrol}{{\color{color_value}\small\sffamily\textbf{UcC}}}
\newcommand{\vfairness}{{\color{color_value}\small\sffamily\textbf{F\&I}}}

\newcommand{\scNew}{{\small\sffamily\textbf{S1}}\xspace}
\newcommand{\scExisting}{{\small\sffamily\textbf{S2}}\xspace}
\newcommand{\scExpert}{{\small\sffamily\textbf{S3}}\xspace}

\newcommand{\scNewA}{{\small\sffamily\textbf{S1}}$_a$\xspace}
\newcommand{\scNewB}{\setulcolor{color_keyword}\ul{{\small\sffamily\textbf{S1}}}$_b$\xspace}
\newcommand{\scExistingA}{{\small\sffamily\textbf{S2}}$_a$\xspace}
\newcommand{\scExistingB}{\setulcolor{color_keyword}\ul{{\small\sffamily\textbf{S2}}}$_b$\xspace}
\newcommand{\scExpertAA}{{\small\sffamily\textbf{S3}}$_{aa}$\xspace}
\newcommand{\scExpertAB}{{\small\sffamily\textbf{S3}}$_{ab}$\xspace}
\newcommand{\scExpertBA}{{\small\sffamily\textbf{S3}}$_{ba}$\xspace}
\newcommand{\scExpertBB}{{\small\sffamily\textbf{S3}}$_{bb}$\xspace}

\definecolor{color_selected}{HTML}{00D1B2}
\definecolor{color_suggested}{HTML}{3E8ED0}
\definecolor{color_network}{HTML}{7A7A7A}
\definecolor{color_keyword}{HTML}{FFE08A}
\definecolor{color_value}{HTML}{7849E5}

\usepackage{soul}
\setul{0.3ex}{0.3ex}
\newcommand{\selected}{\setulcolor{color_selected}\textit{\ul{selected}}\xspace}
\newcommand{\suggested}{\setulcolor{color_suggested}\textit{\ul{suggested}}\xspace}
\newcommand{\network}{\setulcolor{color_network}\textit{\ul{citation network}}\xspace}
\newcommand{\kwords}{\setulcolor{color_keyword}\textit{\ul{keywords}}\xspace}

\newcommand{\us}[2]{\cmarker{#1}\,{\small\sffamily\textbf{US#2}}}

\usepackage{fmtcount} 
\newcommand{\first}{\ordinalnum{1}\xspace}
\newcommand{\second}{\ordinalnum{2}\xspace}

\newcommand{\minuss}{\scalebox{0.9}{\sffamily\textbf{-\,-}}\xspace}
\newcommand{\minus}{\scalebox{0.9}{\sffamily\textbf{-}}\xspace}
\newcommand{\plusss}{\scalebox{0.9}{\sffamily\textbf{+++}}\xspace}
\newcommand{\pluss}{\scalebox{0.9}{\sffamily\textbf{++}}\xspace}
\newcommand{\plus}{\scalebox{0.9}{\sffamily\textbf{+}}\xspace}
\newcommand{\neutral}{\scalebox{0.8}{\sffamily\textbf{o}}\xspace}

\newcommand{\BibTeX}{{\scshape Bib}\TeX\xspace}

\usepackage[normalem]{ulem}
\newcommand{\added}[1]{{#1}}
\newcommand{\deleted}[1]{}

\begin{document}



\firstsection{Introduction}

\maketitle

\label{sec:introduction}

Every scientific work builds on previous work, assembled in a thorough literature search and exploration process. Aside from keyword search, following citation links is a main strategy to discover relevant publications. 
Traditional academic search engines support the process, however, primarily through keyword search. Citations are often included, but only as secondary assets. For instance, in \textit{Google Scholar}~\cite{googlescholar}, citations can be viewed just for a single publication at a time in a list with non-transparent ranking. 
In contrast, targeting a more \textit{exploratory search}~\cite{10.1145/1121949.1121979}, a new generation of literature discovery tools
leverages citation links for suggesting related publications based on selected seed publications. In terms of a visual analytics process~\cite{pirolli2005sensemaking}, such systems  support a foraging loop, going beyond search, but helping build a data collection. Having conceptual predecessors in academia~\cite{10.1145/223904.223913, 10.1145/2851581.2892334, 10.1002/asi.24171}, \textit{CitationGecko}~\cite{citationgecko} was among the first to combine a citation-based recommendation mechanism with a citation network visualization in a cluster and timeline mode. \textit{Litmaps}~\cite{litmaps} also works with a citation network representation in a timeline representation. In contrast, \textit{Connected Papers}~\cite{connectedpapers} relies on a cluster-oriented network visualization mode and encodes temporal information through color. \textit{ResearchRabbit}~\cite{researchrabbit} works with sequences of network visualizations in both cluster- and timeline-oriented modes, and further allows switching to an author-based view of the citation network.
\textit{Inciteful}~\cite{inciteful} includes, aside to publication and author perspectives, also institution and journal views of the network in a more list-oriented interface.
But despite this broad adoption of citation-based search, we still see room for improvement, mainly (i) that it often does not become clear why publications are recommended and which recommendations are most relevant, and (ii) that users lack control to steer the discovery process (a detailed comparison of the tools can be found in the supplemental materials).

In this work, we build on these advances and present \toolname, a citation-based literature search system that is based on explained rankings of suggested publications and visualizations of the citation network. 
Addressing the observed shortcomings, its design has been guided by the values of \textit{transparency and explainability}, \textit{user-centered controls}, and data \textit{fairness and inclusion}. Targeting an audience of researchers and students, these values led to the development of a simple, easy-to-explain recommendation mechanism, which is embedded into a visual interface, subdivided into three main components as showcased in \Cref{fig:teaser}. First, the \selected publications on the left represent the gathered collection of publications and are initially populated with seed publications known beforehand or added through keyword search. Second, on the right, publications are \suggested based on frequent citation links from and to the currently selected publications. Third, below, a \network visualizes the connections between the selected and suggested publications and embeds them into further context (e.g., timeline, keywords, authors). An option to define \kwords---displayed in the header---steers the recommendation process and ranks those publications up that match the desired topics. The interface supports an iterative workflow, where the selected publications are constantly refined to complete a literature collection.

We hence contribute a keyword-controlled citation-based literature search and citation analysis, as well as insights from trying to make fair and transparent recommendations. The discussion of related work in \Cref{sec:related_work} provides a broader overview of similar approaches. As discussed in \Cref{sec:design}, \toolname is designed to support building new collections of literature, updating or assessing existing collections, and identifying experts in a field. Its interface, described in \Cref{sec:interface}, guides through the process of scoping the search, refining the selection of publications, and exploring the citation network. \Cref{sec:implementation} reports on specifics of the implementation such as the data handling and usability. The evaluation, reported in \Cref{sec:eval}, is twofold. In a review of ranking results, we show that the recommendations are reliable and improve through the keyword-based control. Moreover, a user study provides insights into how users apply the tool. On this basis, we can \added{discuss in \Cref{sec:discussion} the applicability of the approach and the implementation of values, as well as} conclude in \Cref{sec:conclusion} that \toolname successfully supports identifying relevant literature.

The developed system is available online\footnote{\scriptsize\url{https://fabian-beck.github.io/pure-suggest/}} and published as open source\footnote{\scriptsize\url{https://github.com/fabian-beck/pure-suggest/}}; this work describes the implementation in version \textit{v0.10.1}. The supplemental materials of the paper contain a video demonstration of the system and detailed materials on the performed evaluation. The work is based on a previous poster presentation at EuroVis 2022~\cite{10.2312/evp.20221110} and has been significantly extended since by, among other features, adding search and filter functionality, refined keyword boosting, keyword and author nodes in the network visualization, and author lists (for details, see the release notes\footnote{\scriptsize\url{https://github.com/fabian-beck/pure-suggest/releases}} since version \textit{v0.2.1}); the poster publication also lacks a discussion of values, user stories, or empirical evaluation. 

\section{Related Work}
\label{sec:related_work}

Going beyond the comparison with similar tools, the following discussion focuses on related research contributions. The collection of referenced works has been created by using \toolname itself. We started from an earlier collection on the topic created with a previous version of \toolname~\cite{10.2312/evp.20221110}, which we updated and broadened. 
An additional keyword search with \textit{Google Scholar} has revealed a three additionally relevant works~\cite{10.1007/978-3-319-22723-817, 10.1145/3383583.3398599, 10.1145/3544548.3580847}.
The final collection is available in the supplemental materials as a session file for \toolname and contains a few more publications than discussed here (e.g., earlier versions of publications). 


\paragraph{Visual Citation-based Literature Discovery} The idea of leveraging the citations of publications for visually discovering related publications has been discussed for some time. In an early work, Mackinlay et al.~\cite{10.1145/223904.223913} visualize citations in butterfly-like lists with a wing each for the outgoing and incoming citations of a publication, which can be interactively navigated through and allows for arranging publications into piles. Some follow-up work has further explored the paper organization and comprehension process~\cite{10.1145/1357054.1357161}\added{, the interactive unfolding of citation networks~\cite{10.1007/978-3-319-22723-817}, and augmenting in-paper citation links~\cite{10.1145/3544548.3580847}}. Most related for us are approaches that focus on search and exploration, such as \textit{PaperQuest}~\cite{10.1145/2851581.2892334}, which presents a list of related papers and explains, like \toolname, the citation-based recommendation relevance score. Citation links can be blended in on demand for individual papers, but the approach does not provide an overview of the citation network. \textit{PaperPoles}~\cite{10.1002/asi.24171} shows groups of citation-based recommended publications and explains them as clusters based on textual similarity. The approach links the suggested publication to keywords, but does not let users choose the keywords to control the search and only partly explains the recommendation score. 
\textit{VisIRR}~\cite{10.1145/3070616} allows switching between content-based, citation-based, and co-author-based recommendations for a selection of publications, and results can be explored as dots in a projected space. However, the scope is wider and recommendations are not highlighted or explained in the visualization.
\textit{RecVis}~\cite{10.1145/3383583.3398599} shows connections of a single selected publication---but not for multiple ones---to recommended similar ones based on a score that integrates various types of similarity.



\paragraph{Analytical Citation Visualization} Not mainly providing publication recommendations, other interfaces target rather higher-level analysis questions. Citation analysis can be the basis for computing meaningful two-dimensional embeddings of publication collections~\cite{10.1109/tvcg.2016.2598667}. Common examples are visualizations of the citation network as node-link diagrams. For instance, in \textit{CiteSpace~II}~\cite{10.1002/asi.20317}, like in \toolname, a cluster view and a timeline view complement each other to visualize citation networks.
\textit{CitNetExplorer}~\cite{10.1016/j.joi.2014.07.006}, as well as \textit{VisualBib}~\cite{10.1016/j.knosys.2019.07.031, 10.1002/asi.24578} and \textit{VisualBib$^{(va)}$}~\cite{10.1109/access.2022.3153027}, arranges node-link citation networks temporally and provides rich navigation and exploration features.
\textit{PaperVis}~\cite{10.1111/j.1467-8659.2011.01921.x} organizes the citation network in layers and assigns publications based on their importance.
\textit{LitSense}~\cite{10.1145/3399715.3399830}, like \toolname, enriches a citation network with topics, but also adds other analytical perspectives.
\textit{Argo Scholar}~\cite{10.1145/3511808.3557177} focuses on manually assembling and sharing of citation networks.
Visual analytics solutions haven been developed around such citation visualizations, which can be combined with automatic processing of the textual document contents~\cite{10.1002/asi.22652, 10.1007/s12650-023-00941-3, 10.1007/s12650-018-0483-5}. 
Other approaches focus on authors, for instance, linking them through co-citations~\cite{10.1016/s0306-4573(98)00068-5, 10.1002/asi.21309} or identifying relevant related authors for a selected publication~\cite{10.1007/s11192-021-03959-2}.
Academic influence can be estimated through citations and put at the center of a visual analysis~\cite{10.1057/palgrave.ivs.9500156, 10.1016/j.jvlc.2018.08.007, 10.1109/access.2019.2932051, 10.1145/2212776.2212796, 10.1145/3132744}. 
Temporal research trends might also be of interest~\cite{10.1145/1056808.1057069, 10.1109/tvcg.2015.2467621}.
\textit{PivotSlice}~\cite{10.1109/tvcg.2013.167} takes publications and their citations as an example for demonstrating an analytical visual querying interface.
Even broader discussions of visualizing citation data can be found in respective survey publications of visualizations for \textit{scientific literature and patents}~\cite[Section 3.2]{10.1109/tvcg.2016.2610422} and for \textit{scholarly data}~\cite{10.1109/access.2018.2815030}. While these examples show that citation data is a rich data source that could inform many use cases, we tried to keep the scope of the tool focused on the specific task of recommending related publications to incrementally assemble a collection. However, we also build on some ideas from the referenced methods, such as representing citation networks as contextualized node-link diagrams.

\paragraph{Visual Literature Search and Exploration} Complementing traditional search interfaces, visual interfaces have been proposed for retrieving, structuring, and presenting literature collections. They visually represent or augment the publications, often using additional data such as keywords, topics, or citations. Regarding retrieval, traditional list-based views can be visually supplemented with, for instance, keywords~\cite{10.1002/asi.24623} or glyph-based citation visualizations~\cite{10.1007/s00799-016-0170-x}.
\textit{PaperCube}~\cite{10.1109/icdim.2009.5356798} and \textit{BEX}~\cite{10.1145/2740908.2742018} provide visualizations to explore citations and other contextual information of a publication.
Keywords and topics can be used to visually guide the search process~\cite{10.1007/s00371-019-01721-7, 10.1109/vis47514.2020.00052}.
Some approaches target specifically the needs of novice researchers~\cite{10.1145/2968220.2968242,10.1109/pacificvis52677.2021.00037}.
Presenting a literature collection, \textit{SurVis}~\cite{10.1109/tvcg.2015.2467757} is a literature browser that visualizes keywords and citation information of a literature collection. 
\textit{VIS Author Profiles}~\cite{10.1109/tvcg.2018.2865022} takes an author-centric perspective and summarizes as an interactive report the publications of the author regarding keywords and collaborations. 
\textit{VIStory}~\cite{10.1007/s12650-020-00688-1} analyzes the figures of publications and organizes them in topic streams.
\textit{ThoughtFlow}~\cite{10.1109/tvcg.2018.2873011} intends to support the ideation of new research projects through topic- and citation-based visualizations that integrate the search and writing process.
Overall, we hence observe a rich variety of approaches supporting diverse literature-related research tasks. Our solution adds to these solutions and fits in between traditional and augmented search solutions, on the one side, and literature exploration tools, on the other.






\begin{table*}[tbp]
	\setlength{\tabcolsep}{3pt}
	\renewcommand{\arraystretch}{1.2}
	\centering
	\caption{Main user stories, grouped by concern (\cmarker{color_selected}~\textit{selected},
 	\cmarker{color_keyword}~\textit{keyword}, 
	\cmarker{color_suggested}~\textit{suggested},
	\cmarker{color_network}~\textit{network}), 
    referencing \textit{scenarios} (\scNew, \scExisting, \scExpert; \syes \textit{fully relevant}, \spartly \textit{partly relevant}) and {\color{color_value}\textit{values}}; user ratings (average value and distribution) on a five-point scale from \cmarker{red}~$-2$ (\minuss) to \cmarker{blue}~$2$ (\pluss) summarize user feedback.
	}
	\label{tab:user_stories}
	\tablefontsize
	\sffamily
	\begin{tabularx}{\textwidth}{@{\extracolsep{4pt}}lXccccp{1cm}}
		\toprule
		\textbf{User Story} &
		\textbf{Description} (\textit{As a researcher or student, \ldots}) &
        \scNew & \scExisting & \scExpert &
        {\color{color_value}\textbf{Values}} &
        \textbf{Rating} \\
		\midrule
        
        \us{color_selected}{1}: \textit{Multi-seed selection} & \textit{I can select publications (known and also unknown through search) as seeds of a new citation-based search, so that I can set the scope of the search.} & \syes & \syes & \syes & \vcontrol & $1.6$ \evalsparklineus{0}{0}{2}{2}{5}\\

        \midrule
        
        \us{color_keyword}{2}: \textit{Keyword-based control} & \textit{I can specify relevant keywords for my current research theme, so that I can steer the suggestions towards more relevant recommendations.} & \syes & \syes & \syes & \vcontrol, (\vfairness) & $1.4$ \evalsparklineus{0}{1}{1}{2}{5} \\ 
        
        \midrule
        
        \us{color_suggested}{3a}: \textit{Suggestion ranking} & \textit{I can see relevant suggestions based on incoming and outgoing citation links in an easy-to-understand ranking, so that I can review the suggestions efficiently and make informed choices.} & \syes & \syes & \spartly & \vtransparancy & $1.3$ \evalsparklineus{0}{0}{0}{7}{2}\\
        \us{color_suggested}{3b}: \textit{Publication character} & \textit{I can judge the character of a publication, so that I am aware of the role of the publication and do not miss potentially underrepresented publications.} & \syes & \syes & \spartly & \vtransparancy, \vfairness & $1.0$ \evalsparklineus{0}{0}{3}{3}{3} \\
        \us{color_suggested}{3c}: \textit{Filter} & \textit{I can filter the suggested publications through metadata, so that I can efficiently retrieve specific publications of interest.} & \syes & \syes & \spartly & \vcontrol & $1.4$ \evalsparklineus{0}{0}{0}{5}{4} \\
        \us{color_suggested}{3d}: \textit{Incremental refinement} & \textit{I can add suggested publications to the selection (or discard them) and receive updated suggestions, so that I can incrementally refine the selection.} & \syes & \syes & \spartly & \vcontrol & $1.9$ \evalsparklineus{0}{0}{0}{1}{8} \\

        \midrule
        
        \us{color_network}{4a}: \textit{Clusters} & \textit{I can identify citation clusters among the selected and suggested publications and contextualize them within the provided keywords, so that the thematic grouping of publications becomes clear.} & \syes & \syes & \syes & \vtransparancy, \vcontrol & $0.9$ \evalsparklineus{0}{0}{3}{3}{3} \\ 
        \us{color_network}{4b}: \textit{Timeline} & \textit{I can inspect the temporal distributions of selected and suggested publications, so that I can judge the history and timeliness of the collection.} & \spartly & \syes & \syes & \vtransparancy & $1.7$ \evalsparklineus{0}{0}{0}{2}{7} \\ 
        \us{color_network}{4c}: \textit{Relevant authors} & \textit{I can see, at a glance, the most relevant authors of the selected publications, so that I can identify the current experts in the area.} & \spartly & \spartly & \syes & \vtransparancy, \vfairness & $1.0$ \evalsparklineus{0}{1}{1}{4}{3} \\
        
		\bottomrule
	\end{tabularx}
	
\end{table*}

\section{Design Considerations}
\label{sec:design}

\toolname was created to make literature search more efficient. As a system being designed for researchers and students of all disciplines and experience levels, it should be generally easy to understand and friendly to use. But even more, it should hold to certain values and ethical standards to prevent unclear results and avoid drawing a biased picture. Linked to these values, we discuss usage scenarios and translate them into specific user stories that define \toolname's main functionality. 

\paragraph{Values}

Providing recommendations comes with responsibility. If suggesting one publication, another is not suggested in the same place. Moreover, users might only trust recommendations if sufficiently explained and want to control the suggestions~\cite{10.17185/duepublico/75905}. We have condensed such considerations into \textit{values} that guided the development of \toolname (marked in {\color{color_value}purple} and referenced throughout the paper).
\textbf{Transparency and Explanation (\vtransparancy)}: Recommender systems only produce value if users can trust the recommendations. Making transparent and explaining recommendations contribute to creating trust~\cite{10.17185/duepublico/75905} and can be supported through visualization~\cite{10.1145/3672276}. We hence strive for a recommendation approach that users can fully understand through the provided visual hints and explanations.
\textbf{User-centered Control (\vcontrol)}: Further, users need to be able to steer the recommendation process~\cite{10.17185/duepublico/75905}. While a basic control mechanism is given through the selecting publications as seeds, we seek further opportunities to enhance the controllability and try to go beyond existing approaches. 
\textbf{Fairness and Inclusion (\vfairness)}: Scientific citations include inherent biases. There are known issues that harm citational justice~\cite{10.1145/3411763.3450389}, such as a gender citation gap~\cite{10.1017/S0020818313000209}. This is not only unfair to the authors of the affected publications, but also inefficient, as relevant research might be overlooked. Whereas we cannot rule out these biases in the citation data we work with, we can try to design the data processing and presentation to mitigate such biases. We strive for an inclusive system that gives relevant, but potentially underrepresented publications a fair chance to be discovered.

\paragraph{Usage Scenarios}
\label{sec:usage_scenarios}
\toolname is designed to support three scenarios that have a collection of publications at the center of the analysis. Users are researchers and students who are focusing on a certain topic, which shall be reflected in the collection.
\textbf{S1 -- Building a new collection:} In this default scenario, the goal is to build a new collection of publications that can be used, for instance, as related work of a specific research paper. Assumptions are that the scope of the research is clear when starting the search, some seed publications are already known or can be retrieved through keyword search. 
\textbf{S2 -- Updating or assessing an existing collection:} Here, a collection already exists, for instance, when a literature search done some time ago needs to be updated or the quality of a collection should be assessed for completeness, for instance, as part of a reviewing or grading process. We have discussed this scenario already in an earlier workshop contribution~\cite{10.5281/zenodo.7123500}.
\textbf{S3 -- Identifying experts through publications:} The third use case does not target the collection of literature itself, but uses a new or an existing collection as a stepping stone for identifying relevant researchers in a field as reviewers, collaborators, speakers, etc. 
Although related to these scenarios, the creation of literature surveys is explicitly \textit{not} supported, as such works typically rely on systematic keyword searches with clearly documented inclusion and exclusion criteria.

\paragraph{User Stories}
As established in software engineering for expressing high-level user requirements, we define the main features of \toolname as user stories (\Cref{tab:user_stories}). We follow the popular \textit{Connextra} sentence pattern \textit{``As a \textless role\textgreater, I can \textless capability\textgreater, so that \textless receive benefit\textgreater.''} Since the system solely targets researchers and students, the prefix \textit{``As a researcher or student''} is the same for all stories. These stories had not been fully elicited upfront, but are the byproduct of an iterative development process (see \Cref{sec:implementation}). 
In \Cref{tab:user_stories}, the user stories are structured and color-coded along the main concerns in the interface, and linked with the previously introduced scenarios. Across all scenarios (\scNew, \scExisting, \scExpert), a core functionality is to select seed publications that define the scope of the respective search session (\us{color_selected}{1}) and to define keywords that steer the search process (\us{color_keyword}{2}). Different features will be necessary to review suggested publications and add them to the selection (\us{color_suggested}{3a--d}), which specifically supports scenarios building and updating literature collections (\scNew, \scExisting), but is only partly relevant for identifying experts (\scExpert; assuming that some relevant literature has been already collected). Finally, a group of user stories (\us{color_network}{4a--c}) refers to the exploration of the citation network, with clusters (\us{color_network}{4a}) being relevant across all scenarios, but a temporal perspective (\us{color_network}{4b}) more relevant for updating collections and identifying experts (\scExisting, \scExpert); inspecting authors (\us{color_network}{4c}) clearly matches the latter (\scExpert).
The user stories will be referenced in the following to describe how they are implemented in \toolname (\Cref{sec:interface}). They also reflect the targeted values, which will be discussed together with the interface design.

\section{\toolname: Interface and Visualization}
\label{sec:interface}

The interface of \toolname is structured into three color-coded main views: \selected, \suggested, and \network (see \Cref{fig:teaser}). This division clearly discerns already selected publications from new publications that are suggested, while the citation network visualizes the relationships among selected and suggested publications. The layout also guides through the search process, as described in the following. 

\subsection{Scope the Search Session}

\begin{figure}
\centering
\includegraphics[width=\columnwidth]{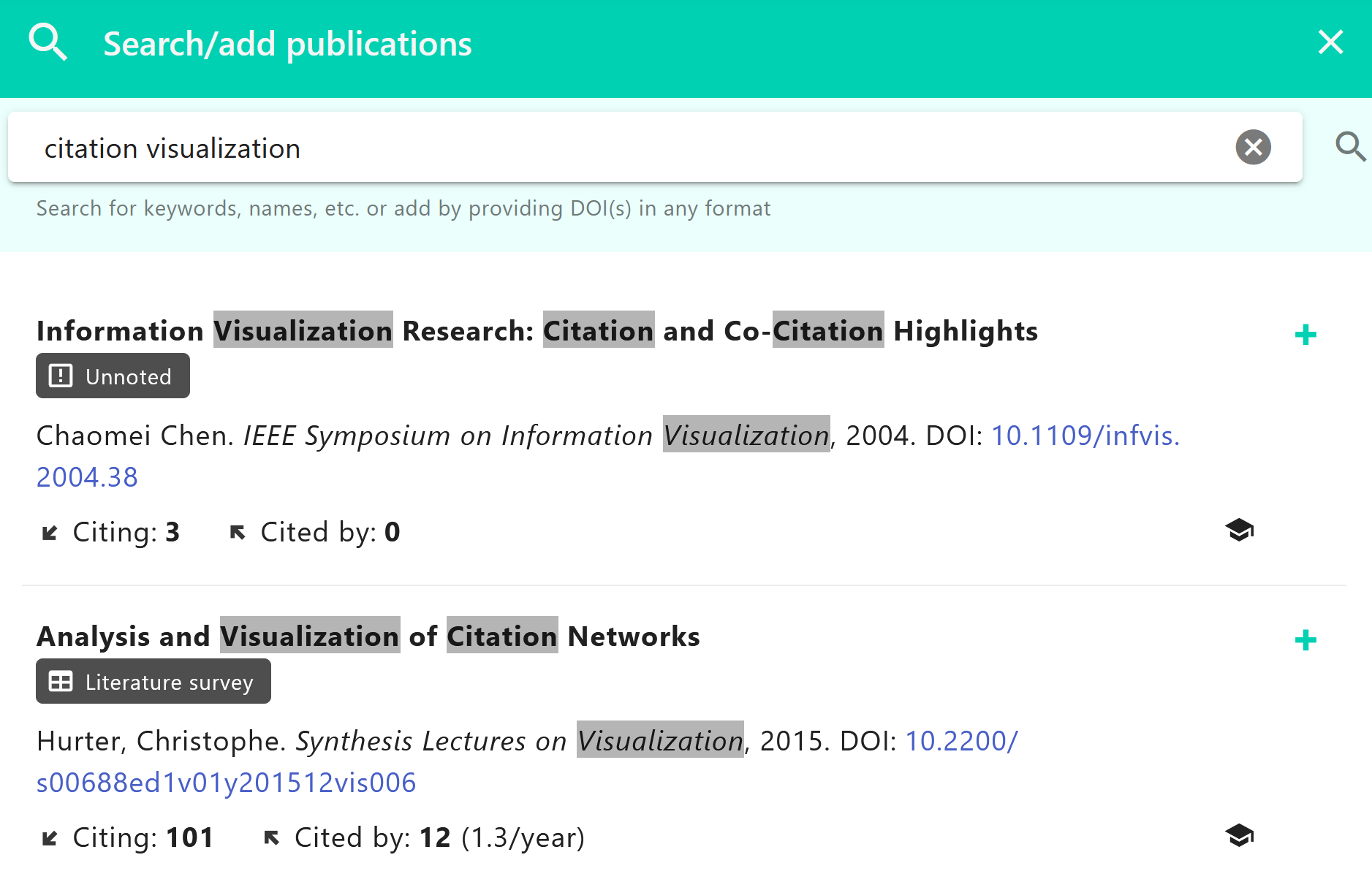}
\caption{Dialog to search for publications based on a query (here: \textit{``citation visualization''}) or to add publications by providing DOIs.}
\label{fig:search}
\end{figure}

Each search session has a scope---the theme of the publications the user intends to find. \toolname uses two main mechanisms to set a scope. First, it is controlled through the \selected publications. Hence, at the start of a  session, a few publications---typically two to five---need to be manually added, which are later constantly refined in the search process. Second, the user can define \kwords to influence the ranking of suggested publications. This cross-cutting aspect of keywords affects all views and has a separate color reserved in the interface.

\paragraph{Add Seed Publications} For initially adding publications as seeds (\us{color_selected}{1a}), \toolname offers a search dialog (\Cref{fig:search}), which can be opened from the \selected publications panel (\inlinegraphic{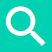} button at the top). It supports two modes, a keyword search and a DOI-based choice of publications. If the user input contains DOIs (\textit{Digital Object Identifiers}), the dialog will load and list the identified publications and users may add all or a subset. Otherwise, a keyword search is performed, and the respective top 20 results are listed. Matches of individual keywords are highlighted through gray background.
These features initilize the user-controlled search process (\vcontrol). After an update (see below), the publications appear in the list of selected publications (\Cref{fig:teaser}, left). Each entry contains the publication title, (abbreviated) authors, and publication year. On click, the entry expands with more metadata appearing (full author list, publication venue, and DOI), as well as citation statistics for outgoing citations (\textit{citing}) and incoming citations (\textit{cited by}) and action buttons (showing the publication's abstract if available; \textit{Google Scholar} search for the publication; \BibTeX export). There is also a publication score shown and further controls, which will be discussed below.

\paragraph{Define Keywords} Addressing user story \us{color_keyword}{2}, the user-defined \kwords allow prioritizing certain publications and boost their ranking if their title matches the keywords (\vcontrol); this might also help thematically matching but potentially underrepresented research to be ranked higher (\vfairness). 
The currently active boost keywords are shown in the header of the interface (see \Cref{fig:teaser}) and can be edited in a text field that opens on click. The syntax of the boost keywords allows defining multiple different keywords separated by commas, and alternative terms for a keyword separated by a vertical line character. For instance, the string \textit{``CIT, VISUAL, MAP, PUBLI|LITERAT''} expresses four keywords, the last one with an alternative. They are written in all capital letters because their definition is case-insensitive. Occurrences of the boost keywords are highlighted across all views, through thick yellow underlining. As exact string matching is used, typically, it is commendable to use a word stem to refer to a group of terms (e.g., \textit{visual} for \textit{visual}, \textit{visually}, \textit{visualization}). 


\subsection{Build the Collection}

At the core of \toolname is a simple, yet effective recommendation approach (see \Cref{sec:eval:quality} for evaluation). It leverages the citation links of the currently selected set of publications to offer suggestions for related publications (\suggested). 
Each publication that receives citations from selected publications or references any of them is ranked. The ranking score $s$ sums outgoing (\emph{citing}, $o$) and incoming (\emph{cited by}, $i$) citation links to or from selected publications. This sum $s_0= (o+i)$ forms a basic approach already. In addition, for steering the recommendations through previously defined \kwords, a degree $b$ may boost the ranking, reflected by the number of matched keywords in the title of the publication. For a substantial boost effect, the final score $s_b = (o + i) \cdot 2^b$ raises $2$ to the power of $b$. The keyword menu allows switching between the base score $s_0$ and the keyword-boosted score $s_b$. The score is also applicable to already \selected publications, where a virtual self-citation is added to the incoming citations. Here, the score acts as a measure of centrality within the citation network (i.e., a variant of degree centrality).

\begin{figure}
\centering
\includegraphics[width=0.7\columnwidth, clip=True, trim=0.1in 0.0in 0.0in 0.0in]{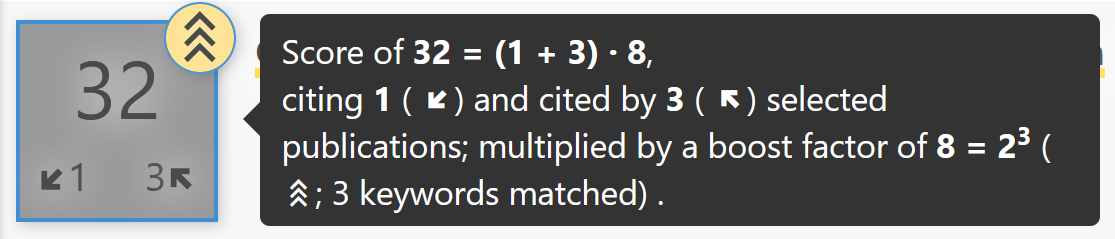}
\caption{Glyph and tooltip explaining the score of a publication.}
\label{fig:glyph}
\end{figure}

\paragraph{Review Ranking of Suggested Publications} 
The \suggested publications are shown, ranked by their score in decreasing order, in the list of publications on the right (\Cref{fig:teaser}; \us{color_suggested}{3a}). Their general layout follows the representation of the selected publications, only the color scheme differs.
Rectangular glyphs visually explain the \textit{score} (\vtransparancy) and hence the ranking order of publications.
Each glyph shows the score ($s_0$ or $s_b$) as a large number, also visually encoded in white to gray background. It lists on the lower left the number of outgoing citation links $o$ (\inlinegraphic{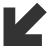}) and on the lower right the number of incoming ones $i$  (\inlinegraphic{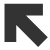}). Chevron icons in yellow circles depict the boost degree, from one chevron in a small circle ($b=1$) to three chevrons in a larger circle ($b\geq3$). As shown in \Cref{fig:glyph}, a tooltip dialog shown on demand explicates this encoding on the concrete example. 

\begin{figure*}
\centering
\includegraphics[width=\textwidth, clip=True, trim=0.2in 0.3in 0.2in 0.3in]{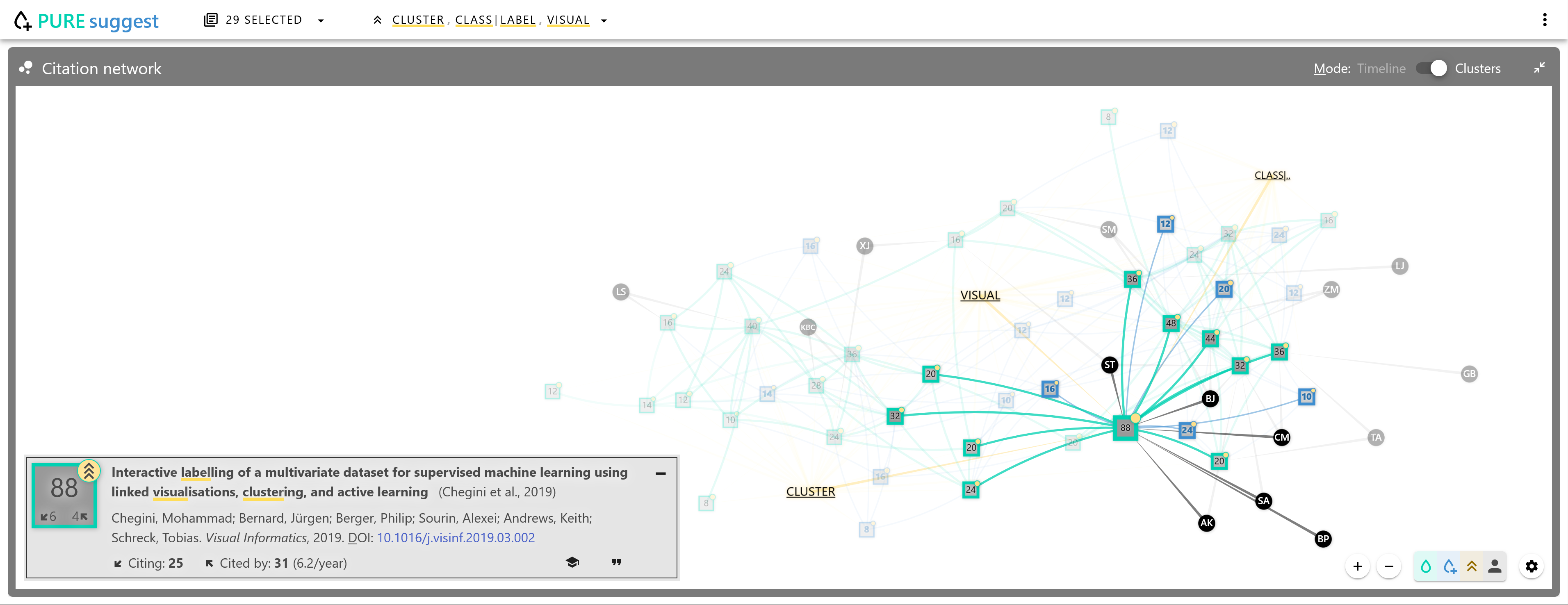}
\caption{The citation network visualization in full-screen, shown in cluster mode in default settings. The collection shows publications related to the \textit{visualization of clustering and classification}, with two clear groups of publications forming around the respective keyword nodes; a central publication linking the two groups is highlighted, with an additional box showing the publication details.}
\label{fig:cluster_vis}
\end{figure*}

\paragraph{Characterize Publications}
\toolname also characterizes and visually tags publications based on their metadata and citation statistics (\us{color_suggested}{3b}). It heuristically classifies them into \textit{highly cited} (more than ten citations per year), \textit{literature survey} (more than 100 outgoing references, or more than 50 and a term such as \textit{survey} in the title), \textit{new} (published at most two calendar years before), and \textit{unnoted} (less than one citation per year). This set of characteristics reflects, on the one hand, what can be realistically estimated with sufficient reliability, and on the other hand, what supports the intended values of \toolname. Generally, hinting at remarkable features such as many incoming citations (\textit{highly cited}) or outgoing references (\textit{literature survey}) supports transparency (\vtransparancy). Moreover, the characteristics \textit{new} and \textit{unnoted} highlight publications that have not yet received considerable attention  (\vfairness).

\paragraph{Filter Suggested Publications} For targeting the search, different filter options allow restricting the list of \suggested publications (\us{color_suggested}{3c}). On activating the filter switch in the upper-right corner of the interface (\Cref{fig:teaser}; \inlinegraphic{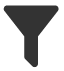}), a filter panel appears at the top of the list.
\linegraphicdouble{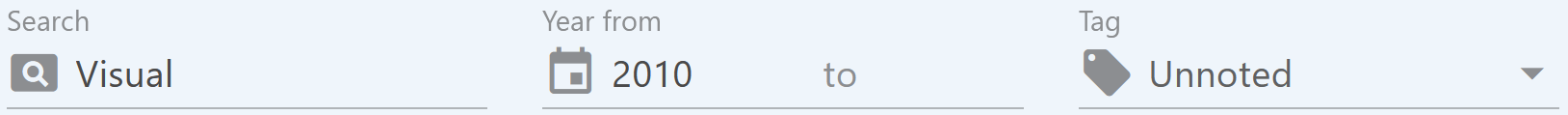}
Available filters support searching for strings in publication titles, to restrict the publication year, and to select one of the characteristics tags. 
When setting any of the fields, the recommendations immediately update, together with the statistics at the top of the panel.
This further contributes to giving users control over the recommendations shown (\vcontrol) and supports use cases like looking for recent survey articles.

\paragraph{Refine the Selection} A key step in the search process is then deciding which of the suggested publications to include in the selection, or to exclude (\us{color_suggested}{3d}). On the right side of each \suggested publication, two buttons support the inclusion \inlinegraphic{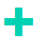} and exclusion \inlinegraphic{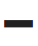}.
Whenever clicking the respective button, however, the publication is only marked to be included or excluded. Changes get only effective when an update is manually triggered. As loading of the new suggestions typically takes a few seconds, this avoids frequent interruptions in the users' workflow. But more importantly, it does not disrupt a systematic inspection of the top suggestions, which might get reordered on update. Marking publications that were not yet focused as \textit{unread} (light blue background; blue font) further supports following which publications were already inspected. These features contribute to providing control over the search process (\vcontrol)---the citations from and to the newly selected publications lead to a refined scope and further recommendations.

\subsection{Explore the Citation Network}

The list of suggested publications allows the inspection of individual publications but lacks context for the publications and an overview of their interconnections. Addressing this, the \network (see \Cref{fig:teaser}, bottom, and \Cref{fig:cluster_vis}) complements a visualization that focuses on integrating the available data. It displays publications as rectangular nodes in the respective colors for \selected and \suggested, as well as includes nodes for \kwords (text labels) and authors (black circles). Buttons allow blending in and out each type of node. The publication nodes are connected through citations to or from selected publications; their direction is encoded through curved links and follow a counterclockwise arc. Keyword nodes are linked to the publications they match, author nodes to the selected publications they contributed to; both are drawn as straight, tapered links. Settings control the number of displayed suggested publications and authors.
 
\paragraph*{Inspect Clusters} The default layout of the network, as shown in \Cref{fig:cluster_vis}, supports identifying clusters of publications (\us{color_network}{4a}). The layout builds on the force-directed model of \textit{D3js}, tailored as follows. While suggested publications are more volatile, we assume the \selected publications to be usually more stable. As such, the layout of these nodes and the links between them can form a backbone structure of the graph. To this end, the targeted length is longer for these links, and they are drawn as relatively opaque lines. While this link length decreases with more publications being selected, the force that pushes apart all nodes increases---this warrants that clusters of publications are clearly identifiable at different stages of the search process and suggested publications can be seen in their context (\vtransparancy). Additionally, \kwords nodes indirectly add semantics to the connected clusters. They can also act as anchors to form a map-like layout. Whereas other nodes are not interactively movable, the users can drag them to an intended position (sticky). Arranging the main keywords provides stable semantic regions in the layout even when selections grow and suggestions change. The users receive control over the layout (\vcontrol), but without overwhelming them with the responsibility to arrange many nodes or getting involved with technical details (e.g., the configuration of forces). 

\paragraph*{View Temporal Trends} For a temporal perspective (\us{color_network}{4b}), as an alternative to the cluster mode, the nodes can also be arranged on a timeline. While publication nodes appear at a specific position on the x-axis based on their publication year, the placement on the y-axis still follows the force-directed model described above. Keyword nodes are also fixated on the x-axis to the right of the timeline, but float vertically according to the force simulation. In contrast, author nodes can move on both spatial dimensions, according to the links to connected publications and a general tendency to the center year of the author's publication span (here: span between first and last selected publication). This mode provides additional temporal contextualization of the suggested publications (\vtransparancy).

\paragraph{Identify Experts}

\begin{figure}
\centering
\includegraphics[width=\columnwidth, clip=True, trim=0.0in 1.0in 0.0in 0.0in]{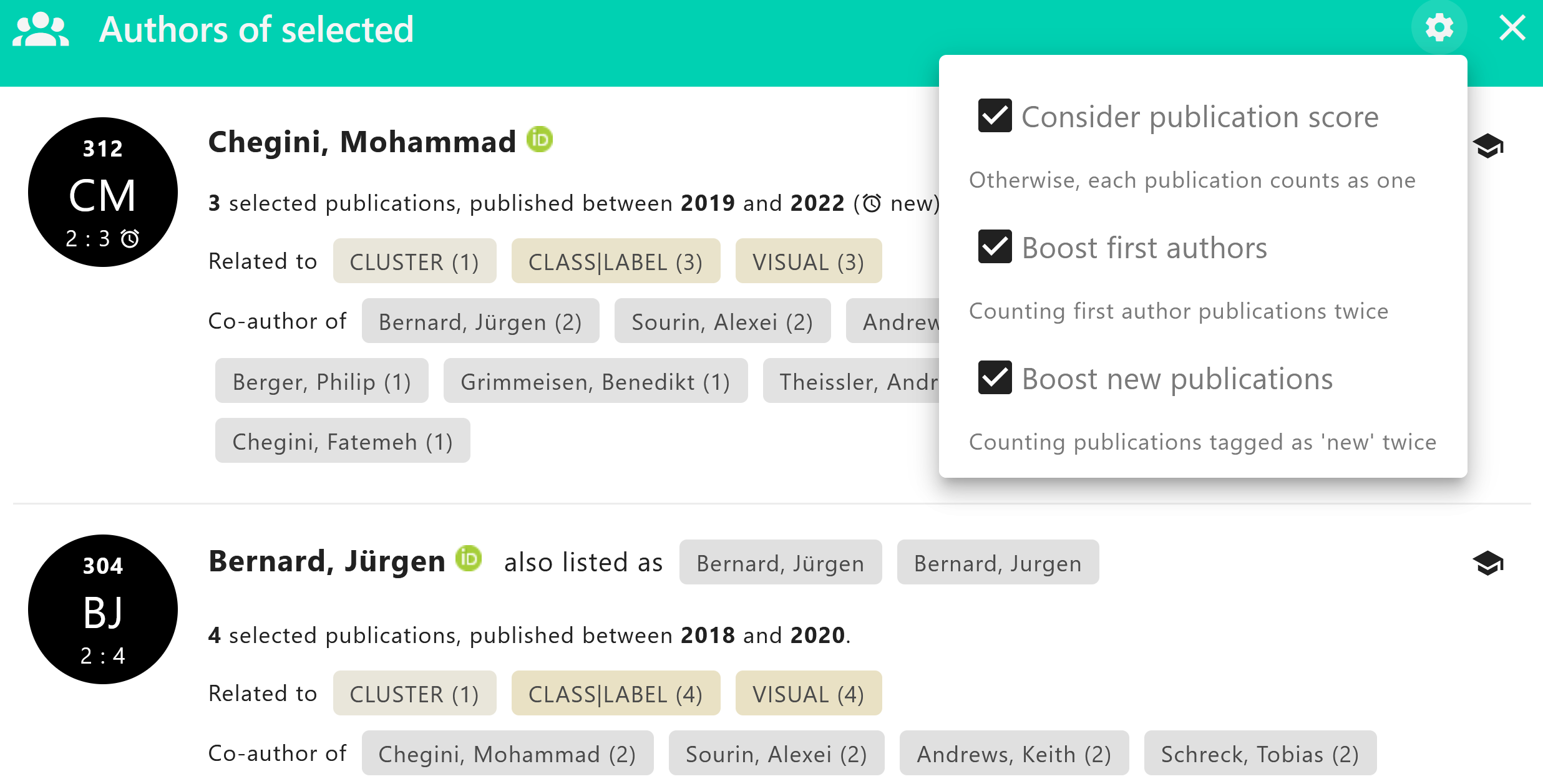}
\caption{Ranked list of authors of selected publications (same as in \Cref{fig:cluster_vis}). For each author, a glyph shows a score together with the author's initials and statistics; the summary below the author name characterizes the co-authored publications in terms of number, year range, matched keywords, and co-authors. A settings menu allows configuring the score.}
\label{fig:authors}
\end{figure}

The list of authors of the \selected publications can be inspected in a modal dialog (button \inlinegraphic{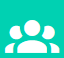} at the top of selected publications). The author list as shown in \Cref{fig:authors} is ordered by a score that relies on the authors' contributions to the selected publications. By default, the score adds the publication scores of the co-authored selected publications, boosting first-author and recent publications by a factor of two each. This intends to improve career-stage diversity and acknowledges higher contributions of first authors (\vfairness). Depending on the exact use case, however, users might want to configure the ranking, which is possible through a settings panel (\Cref{fig:authors}, top-right). For each author in the list, the score is visually explained in a glyph (a tooltip, again, describes details). Next to the glyph, more information related to the author is shown, like the \kwords and co-authors. In the \network, only the top-ranked authors are shown as nodes, but the number can be interactively changed. Hovering an author node shows related publications and a tooltip with a summary of the author information. Clicking on a node opens the author dialog and highlights the author. Being able to see the main authors in the network contributes to transparency (\vtransparancy)---for instance, one might perceive a cohesive cluster of publications differently if knowing that all contained publications are authored by the same team.

\section{Implementation}
\label{sec:implementation}

Up to now, \toolname has undergone 20 releases spread over more than two years. 
During the whole development history, the key mechanisms and scope have been stable, from the beginning supporting scenario \scNew. Although scenarios \scExisting and \scExpert have been added later, they have substantially overlapping user stories (see \Cref{tab:user_stories}).

\paragraph{Data Processing} \toolname relies on openly available citation data. Publicly accessible APIs by \textit{OpenCitations} and \textit{Crossref} serve as data sources, the first listing incoming and outgoing citations per publication, the latter providing more detailed metadata. Furthermore, the query-based search (see \Cref{fig:search}) is supported through requests to the \textit{Crossref} API. DOIs are used as publication identifiers throughout the system, which align with their usage in \textit{OpenCitations} and \textit{Crossref}, however, limits the system to DOI-referenced publications. The data from both sources is merged. In case of missing or faulty metadata, \toolname tries to reconstruct or correct it, for instance, guessing a missing publication year from a DOI, correcting capitalization, and fixing encodings of special characters. Still some missing values and ill encodings cannot be resolved, most severely, missing references (which \toolname indicates in red font in the citation statistics of the publication).
Unfortunately, the availability of abstracts is still relatively limited---this is why we had to restrict all features related to keywords to paper titles.
Moreover, for publications cited 1\,000 times or more, citations cannot be considered because of long request times. However, references to extremely frequently cited publications might not carry much meaning (e.g., the reference to a standard method).

\paragraph{Name Disambiguation} For providing an overview of authors, matching different versions of the same author name is necessary to provide results of acceptable quality. As the same author might be listed with a different spelling of their name, we apply a simple, yet effective name disambiguation. We use an identifier derived from the author name (unify special characters and transform to lower case) and merge all authors having the same unified name, or the same ORCID (\textit{Open Researcher and Contributor ID}; unfortunately, only available for some publications). Additionally, we merge authors with abbreviated names if there is an author where the matching name is not abbreviated. Such preprocessing is especially important to avoid unfair splitting of authors (\vfairness), for instance, from countries where special characters are often contained in the name, or of authors who have changed their name (e.g., through marriage, with matching possible through ORCIDs). 

\paragraph{Architecture and Data Loading} \toolname mainly consists of a web-based front-end implemented using \textit{vue.js} as a single page application framework, \textit{vuetify} for user interface components, and \textit{d3.js} as a visualization library. The front-end implements the interface and includes also the computations of recommendations. In contrast, the back-end, is lightweight: Aside from orchestrating calls to the public APIs of \textit{OpenCitations} and \textit{Crossref}, it only merges the data as described above and caches it. 
While all cited and citing publications are considered for the suggestions, only the top 50 most cited and citing publications are loaded with metadata to the interface to limit loading times caused by calls to the public APIs (keyword boost factors can only be considered later, when the metadata of the suggestions is loaded). Nevertheless, users can choose to load more through a button~\inlinegraphic{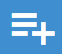}. For similar performance reasons, the search dialog only shows the top 20 results, but this appeared sufficient in all tested sessions.
Pre-fetching is applied to reduce waiting times, already loading the next 50 suggested publications in the background---they might soon be requested by the user or get ranked up. All loaded publications are also locally cached in the browser to avoid redundant requests. 

\begin{figure}
\centering
\fgraphic{height=2.6cm}{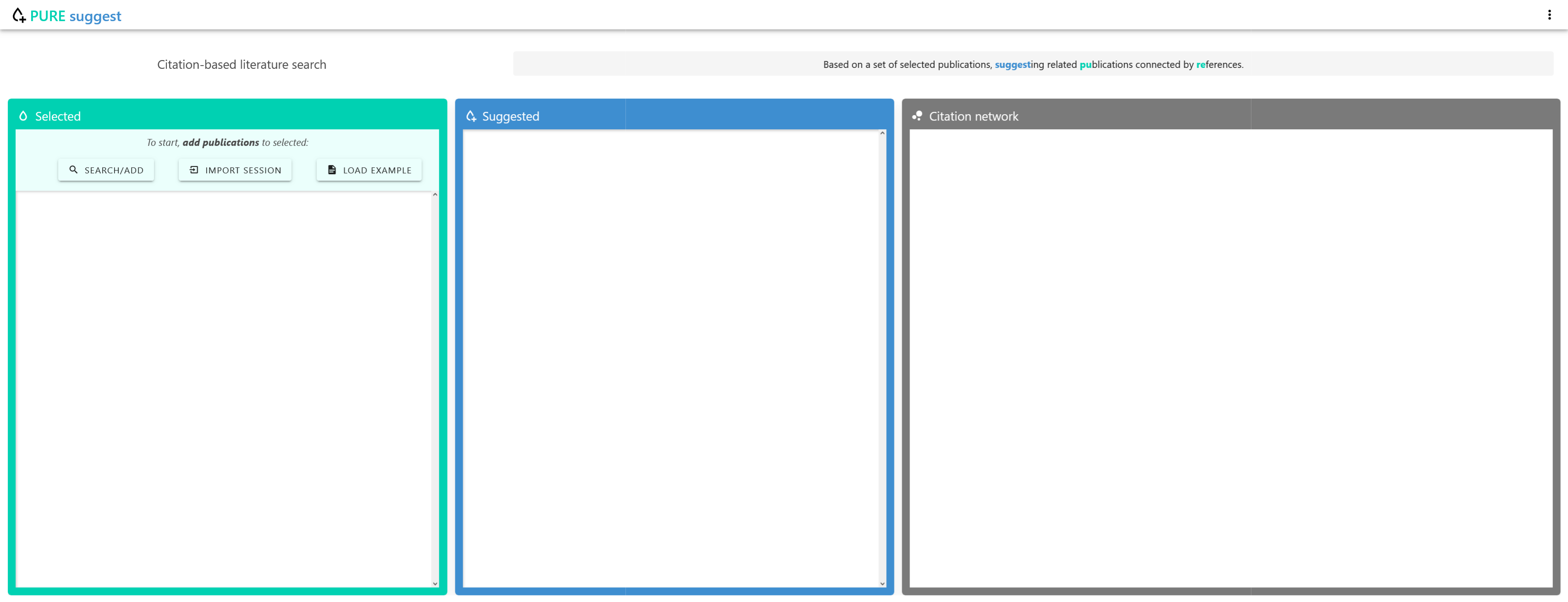}
\fgraphic{height=2.6cm}{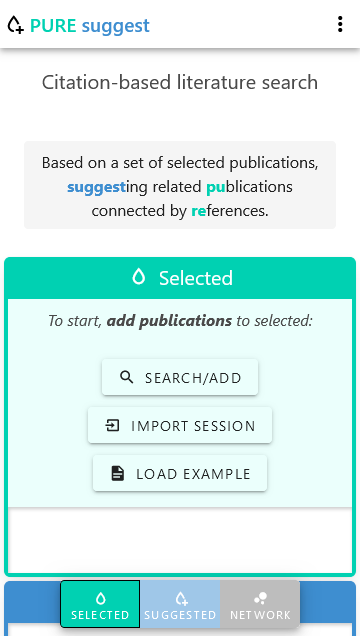}
\caption{Alternative user interface layouts for different screen sizes: (left)~ultra-wide, (right)~mobile; empty selections as shown on start-up.}
\label{fig:ui_layout}
\end{figure}

\paragraph{User Experience} To support flexible use, we made the interface responsive to screen sizes. On common desktop monitor resolutions, the three views are arranged as shown in \Cref{fig:teaser}; on wide screens, however, they all shown side-by-side to use the vertical space, while on mobile screens, they are listed from top to bottom (see \Cref{fig:ui_layout}). Alternatively, on desktop screens, the citation network can be shown enlarged with details of a highlighted publication shown in the left corner of the diagram (see \Cref{fig:cluster_vis}).
On start-up, all three views are empty, but show a brief introduction and options to initialize the search.
When data is loaded, various highlighting interactions connect the views; for instance, clicking a publication node in the citation network highlights the node in the list of selected or suggested publications, as well as hints at publications connected through citations across all views (see video or interactive tool).
For experienced users, keyboard controls allow the quick navigation in the main view (see keyboard controls menu page).
Finally, a session can be saved (and imported again later) as a JSON file or exported as a \BibTeX file.

\section{Evaluation}
\label{sec:eval}

The evaluation is twofold, to provide, on the one hand, estimates of the effectiveness of the recommendation approach and, on the other hand, insights into realistic user behavior. First, the core suggestion mechanism is tested, and it is shown that keyword-based and other steering can improve the publication recommendations and make author rankings more diverse. Second, user feedback and interaction logs are evaluated to confirm the applicability of the tool and to obtain insights into how the system creates value for users. 

\subsection{Recommendation Quality}
\label{sec:eval:quality}

As an evaluation of the result quality, we systematically test if relevant publications are recommended, with and without using \kwords. Simulating scenarios \scNew and \scExisting, we chose different starting points and rated the top ten suggested publications into \textit{fitting} ($1.0$), \textit{partly fitting} ($0.5$), and \textit{non-fitting} ($0.0$); then, summing up the individual fit scores results in an overall fit score between $0.0$ (worst fit) and $10.0$ (best fit). Similarly, we evaluated the ranked authors in scenario \scExpert, also testing different ranking settings. We used six different topics from information visualization and visual analytics as examples, covering a diversity of topics from evaluation of standard diagrams to visualization for machine learning. The decision to focus only on the visualization community was motivated by the observation that the bibliographic data sources (see \Cref{sec:implementation}) well cover the field, and because we are able to judge the quality of suggestions. While the results are summarized in \Cref{tab:quality:onetwo} (\scNew, \scExisting) and \Cref{tab:quality:expert} (\scExpert), a more detailed documentation of the results is part of the supplemental materials. Aside to fit scores, to contribute to the discussion of fairness of recommendations (\vfairness), the tables report on numbers of \textit{unnoted} publications (i.e., works with minimal number of citations, potentially underrepresented), as well as measures for diversity and early career status, respectively.

\begin{table}[tbp]
	\setlength{\tabcolsep}{3pt}
	\renewcommand{\arraystretch}{1.4}
	\centering
	\caption{\textbf{Publication rankings} (\textbf{S1}, \textbf{S2}) in conditions $a$: without keyword boost; $b$:~with keyword boost; average summed \textit{fit score} with a fit value distribution from \cmarker{red}~\textit{non-fitting} to \cmarker{blue}~\textit{fitting}; average number of \textit{unnoted} among the top ten suggested publications.}
	\label{tab:quality:onetwo}
	\tablefontsize
	\sffamily
	\begin{tabularx}{\columnwidth}{@{\extracolsep{4pt}}lCC|lCC}
		\toprule
         
		& \textbf{Fit score}
		& \textbf{Unnoted}
        &  
		& \textbf{Fit score}
		& \textbf{Unnoted}
        \\ 
		\midrule
        \scNewA & \ccell{7.3} \evalsparkline{5}{23}{32} & \ccell{5.7} & 
        \scExistingA & \ccell{8.1} \evalsparkline{4}{15}{41} & \ccell{6.3}
        \\
        \scNewB & \ccell{8.7} \evalsparkline{0}{16}{44} & \ccell{5.3} & 
        \scExistingB & \ccell{8.6} \evalsparkline{0}{17}{43} & \ccell{6.3} 
        \\
		\bottomrule
	\end{tabularx}
\end{table}

\paragraph*{Publication Ranking} \textit{Procedure:} Creating new collections~(\scNew), we seeded each simulated session with three meaningful selected publications and came up with three to six \kwords (defining up to four alternatives per keyword). Then, we rated the top ten suggested publications without considering keywords (\scNewA) and with using the keywords for boosting (\scNewB). 
For updating or assessing an existing collection~(\scExisting), we extended the previously selected publications employing the defined keywords---depending on the breadth of the topic---to $15$ to $78$ publications, only considering publications being published by 2021 at latest. Then, we set a filter to \textit{new} (2022 and younger) and checked the top ten suggested publications of those newer publications (loading the first 200 suggested publications), again without (\scExistingA) and with keyword boosting (\scExistingB).
\textit{Results:} The averaged fit scores in \Cref{tab:quality:onetwo} show high recommendation quality in all conditions ($[7.3, 8.7]$). In \scNew, the keyword-boosted ranking clearly outperforms the ranking without considering the keywords (\scNewA: $7.3$; \scNewB: 8.7). Across the six tested topics, the better performance of the keyword-boosted approach is consistent ($5 \times \textit{better}$, $1 \times$ \textit{equal}; see supplemental materials). In \scExisting, the effect is similar but not as strong (\scExistingA: $8.1$; \scExistingB: $8.6$) and as consistent ($4 \times$\textit{better}, $1 \times$ \textit{equal}, $1 \times$ \textit{worse}). \Cref{tab:quality:onetwo} also lists the average number of \textit{unnoted} publications in the top-ranked suggested publications. The numbers are generally high, however, the keyword-boosted ranking does not list more \textit{unnoted} works, but slightly less (\scNewA: $5.7$; \scNewB: $5.3$) or comparable (\scExistingA: $6.3$; \scExistingB: $6.3$). 

\paragraph*{Author Ranking} \textit{Procedure:} For the identification of experts (\scExpert), we imagined finding potential speakers for a lecture series on the respective topic. We extended the lists of publications from scenario \scExisting, also including newer works now, which resulted in $22$ to $84$ publications. As recommendations, we studied the top ten authors of the selected papers. In a $2 \times 2$ pattern, we tested four ranking conditions, which can be set in the author dialog (see \Cref{fig:authors}):  the number of co-authored selected publications (\scExpertAA; no checkbox active), weighted by the publication score in its default configuration including \kwords boosting (\scExpertAB; first checkbox active), boosting \textit{new} and \textit{first-author} contributions (\scExpertBA; second and third checkbox active), and both additional factors together (\scExpertBB; all checkboxes active). Aside to a topic-wise fit given through the selected papers, we required the authors to be actively publishing on at least related topics. We further recorded whether the author is at an early stage of their career (actively publishing since maximum ten years, with the first publication in 2014 or later). To measure how diverse the authors are regarding their academic network, we counted those authors who did not co-author any selected publications with any of the authors listed above them. 
\textit{Results:} As summarized in \Cref{tab:quality:expert}, the general fit scores are high in all conditions ($[8.0, 8.8]$), but best using the simplest scores (\scExpertAA: $8.8$). While first surprising to us, further analysis clarified that a risk is taken in the other conditions: Authors of a single selected paper are rated higher through the publication scores or by boosting first authors and new publications. While considering the publication score seems to have disadvantages (\scExpertAB lower than \scExpertAA in all measures; \scExpertBB lower than \scExpertBA in all measures), boosting first authors and new publications (\scExpertBA) performs best regarding co-author \textit{diversity} ($6.7$) and \textit{early} career researchers ($3.3$), and decently regarding the overall fit ($8.3$).

\begin{table}[tbp]
	\setlength{\tabcolsep}{3pt}
	\renewcommand{\arraystretch}{1.4}
	\centering
	\caption{\textbf{Author ranking} (\textbf{S3}) in conditions $aa$: without factors; $ab$:~weighted by publication score; $ba$: boost factors for first authors and new publications; $bb$: all factors from $ab$ and $ba$; average summed \textit{fit score} with a fit value distribution from \cmarker{red}~\textit{non-fitting} to \cmarker{blue}~\textit{fitting} and average numbers of authors contributing to co-author \textit{diversity} and at \textit{early} stages of their careers among the top ten authors.}
	\label{tab:quality:expert}
	\tablefontsize
	\sffamily
	\begin{tabularx}{\columnwidth}{@{\extracolsep{4pt}}lcCC|lcCC}
		\toprule
        
		\textbf{} 
		& \textbf{Fit score}
		& \textbf{Divers.}
        & \textbf{Early} 
        & \textbf{} 
		& \textbf{Fit score}
		& \textbf{Divers.}
        & \textbf{Early} 
        \\ 
		\midrule
        \scExpertAA & \ccell{8.8} \evalsparkline{1}{12}{47} & \ccell{4.2} & \ccell{1.8} &
        \scExpertAB & \ccell{8.0} \evalsparkline{5}{14}{41} & \ccell{3.0} & \ccell{1.3}
        \\
        \scExpertBA & \ccell{8.3} \evalsparkline{4}{13}{43} & \ccell{6.7} & \ccell{3.3} &
        \scExpertBB & \ccell{8.2} \evalsparkline{3}{16}{41} & \ccell{4.7} & \ccell{2.2}
        \\
		\bottomrule
	\end{tabularx}
\end{table}

\paragraph*{Summary and Reflection} Across all considered scenarios and tested conditions, \toolname makes relevant recommendations, to a level where almost every of the top suggestions is worth considering (\textit{fitting} or \textit{partly fitting}). The addition of meaningful \kwords generally improves the results, most notably when starting to build a new collection (\scNew). When updating an existing collection (\scExisting), the quality improvement is not as clear; however, in any case, the advantage remains of steering the recommendations in a user-defined way (not directly tested in the study). While a positive effect of keywords boosting on the number of \textit{unnoted} publications could not be observed (\scNew, \scExisting), the overall ratio of \textit{unnoted}---and mostly fitting---publications is high. Identifying experts (\scExpert), boost factors for first authors and newly published works lead to recommendations of clearly more diverse sets of authors and a substantially higher ratio of early career researchers. This comes only at the cost of slightly less fitting recommendations. 

\subsection{User Study}

\begin{figure*}
\centering
\includegraphics[width=\textwidth, clip=True, trim=0.0in 5.6in 0.0in 0.0in]{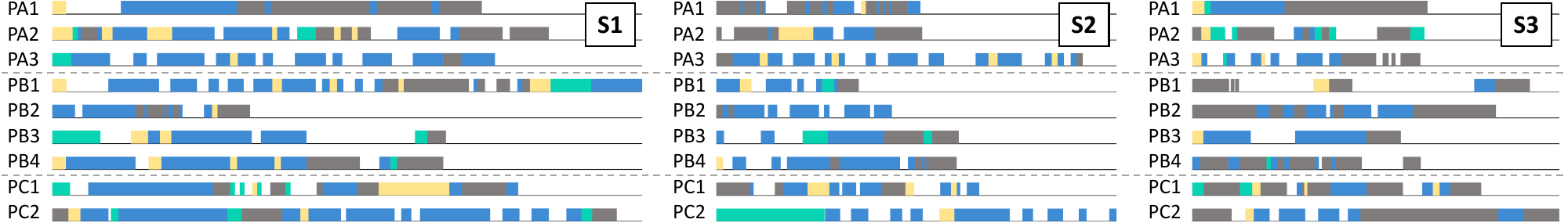} 
\caption{Comparing the user sessions across all participants (rows) and recorded tasks (columns) on a timeline marking periods of substantial usage activity, with the longest session duration (\textit{mm:ss}) of \scNew: 17:40; \scExisting: 12:00; \scExpert: 11:00; color codes refer to the groups of user stories in \Cref{tab:user_stories}, with activities related to \cmarker{color_selected} selected publications, \cmarker{color_suggested}~suggested publications, \cmarker{color_network} the citation network and authors, as well as \cmarker{color_keyword}~keywords. }
\label{fig:sessions}
\end{figure*}

For a more general assessment of \toolname, we performed a user study with nine participants covering all three scenarios (\scNew, \scExisting, and \scExpert). The tasks were linked to the field of information visualization and visual analytics. This allowed us to define meaningful session scopes, as well as to recruit sufficiently experienced participants from our professional network. Moreover, users with expertise in visualization could give potentially more informed judgments. We decided against contrasting \toolname with other tools, as differences in visualization and interaction design appeared too big to allow a direct, meaningful interpretation of potentially observed task performance differences. 
Instead, the goal was an explorative analysis of realistic user behavior, which we recorded through logged interactions with the interface. The recorded events then could be loaded into a timeline-based event visualization, which we developed specifically for the purpose of analyzing the user behavior in this study. The user activities were annotated aside viewing recorded videos of the sessions, and we present a summary of the annotated activities in \Cref{fig:sessions}. We complemented the session recordings with user feedback from a questionnaire.
In the following, we briefly report the study procedure and summarize the findings. A more detailed documentation of used study materials and user feedback is available in the supplemental materials, as well as a documentation of the developed timeline-based visualization approach and respective visualizations of all recorded sessions.

\paragraph{Study Design}
\label{sec:evalu:usage_study}

The study was designed to last about 60 minutes, with an experimenter present and taking notes. The study took place remotely via Zoom, or co-located with Zoom used for screen sharing and recording. Participants used their PCs for the study, and we made sure that they chose a browser window of sufficient size for the standard layout of the interface (see \Cref{fig:teaser}). After being informed about the study conditions and consenting to them, the participants opened \toolname. First, to introduce \toolname, the experimenter guided the users through the main features step by step and provided explanations, similar as presented in \Cref{sec:design}. Then, the main part of the study started, and we activated screen recording. We phrased a task for each scenario and provided a visualization topic for it. The tasks related to \textit{completing a literature collection for a given set of seed publications} (\scNew), \textit{updating an outdated literature collection with the newest approaches} (\scExisting), and \textit{finding speakers for a lecture series}~(\scExpert).
The topics matched the first three topics of the above recommendation quality analysis and were called \textit{T1: Personal Sports and Health Visualization}, \textit{T2: Treemaps}, and \textit{T3: Visualization for Classification and Clustering}.
Each participant worked through all topics, and we rotated the topics with each participant so that different topics were assigned to different tasks (scenarios).
We always kept the sequence of scenarios (\scNew $\rightarrow$ \scExisting $\rightarrow$ \scExpert), as they were intended to reflect different levels, from basic to more advanced.
Study participants were not asked to speak their thoughts aloud always, but with the experimenter present, they often deliberately commented on important actions they took. Participants were allowed to ask questions anytime---e.g., about a task or a certain tool feature---, and vice versa, the experimenter gave occasional advice in cases where participants were stuck, could not find a feature, or partly misunderstood a task.
Finally, after completing the three tasks, we stopped the recording and the participants were asked to fill a questionnaire, in which we inquired about their previous experience in relevant areas, their judgement about the system in general and specific to certain features, as well as the usage strategies they applied.

\paragraph{Participants}
Recruited to cover diverse levels of experience, the nine participants were active students and researchers. We assign them to three levels, with three having a PhD as highest \textit{completed} degree \added{(PA1--PA3)}, four a Master \added{(PB1--PB4)}, and two a Bachelor \added{(PC1--PC2)}. All participants had at least some experience in information visualization and visual analytics, but the majority were experienced in the field ($1 \times$\plus; $4 \times$\pluss; $4 \times$\plusss; nobody answered \neutral). We see a similar distribution of experience in academic literature search ($2 \times$\plus; $4 \times$\pluss; $3 \times$\plusss; nobody answered \neutral). While five participants had not used \toolname before (\neutral), four  answered to have slight experience with it (\plus; nobody answered \pluss or \plusss). Typical tools that participants usually used for literature search and management of references (listing tools mentioned at least twice) were \textit{Google Scholar} ($8 \times$), \textit{Zotero} ($5 \times$), \textit{ResearchGate} ($2\times$) and \textit{Mendeley} ($2\times$). 

\paragraph{User Feedback}
To provide context for the following qualitative analysis, we first summarize the numeric user ratings (as the numbers of participants are small for any quantitative evaluation, however, these results are preliminary and should not be interpreted independent of respective qualitative results, e.g., from the questionnaire).
Ratings of the individual user stories showed large agreement that \toolname supports the stories as intended. Providing with the stories as phrased in \Cref{tab:user_stories}, the participants rated them on a five-point scale from \minuss ($-2$) to \pluss ($2$). The ratings, as also summarized in \Cref{tab:user_stories}, are mostly very positive ($40\times$\pluss) and positive ($29\times$\plus), with only a few neutral ratings ($10 \times$\neutral) and even fewer negative ratings ($2\times$\minus; $0 \times$\minuss). However, there are differences between the stories. We almost only see positive replies with average ratings of $1.3$ or higher for stories related to \textit{multi-seed selection} (\us{color_selected}{1}), \textit{keyword-based control} (\us{color_keyword}{2}), \textit{suggestion ranking} (\us{color_suggested}{3a}), \textit{filter} (\us{color_suggested}{3c}), \textit{incremental refinement} (\us{color_suggested}{3d}), and \textit{timeline} (\us{color_network}{4b}). Inspecting the textual feedback, we find some explanations for these ratings. Repeatedly mentioned as positive was the iterative recommendation approach ($6\times$; \us{color_selected}{1}, \us{color_keyword}{2}, \us{color_suggested}{3a}, \us{color_suggested}{3d}) and the timeline ($5\times$; \us{color_network}{4b}).
In contrast, three stories are rated slightly less positively (average ratings around $1.0$) and get reflected in critical comments.
Regarding judging the \textit{publication character} (\us{color_suggested}{3b}), not analyzing publication abstracts, as mentioned by \added{PB2}\deleted{one participant}, could partly explain the lower score. Related to \textit{clusters} (\us{color_network}{4a}), \added{PB1}\deleted{one participant} found the network to be overwhelming. 
Finally, to \textit{author}-related features (\us{color_network}{4c}), we can link statements by four participants \added{(one unclear)} including requests for a better search for authors \added{(PB1)}, an improved linking of authors to their publications \added{(PA2)}, and easier discoverability of the author ranking \added{(PB4)}.
Other suggestions for improvement related to 
allowing different prioritization of \kwords~\added{(PC1)} and a more accurate and quicker search feature \added{(PA3)}.
Comparing the three scenarios, participants rated higher updating or assessing an existing collection (\scExisting; $4 \times$\first, $4 \times$ \second) and building a new collection (\scNew; $4 \times$\first, $2 \times$ \second) than the identification of experts (\scExpert; $1 \times$\first, $3 \times$ \second).

\paragraph{User Activity Annotation} For investigating the strategies that users applied for searching, we studied user activities across the three scenarios, as shown in \Cref{fig:sessions}. The visualization encodes user activities in color according to the groups of user stories described in \Cref{tab:user_stories}. To create this annotation, we visually analyzed the interaction logs on a timeline visualization (see supplemental materials) and manually labelled the activities. Whenever a user activity was no clear from the log data only, we manually checked the video for assigning the best matching group. White areas mark activities mostly spent outside the interface (clarifying the task description, opening a paper in another tab, etc.). Overall, though varying in depth, all tasks were solved successfully by all participants; the only exception was in scenario \scExpert of the first participant (\added{PB1}), who partly misunderstood the task (we made sure to explain the task better in all follow-up sessions). 

\paragraph{Search Strategies} 
Across all scenarios, we observe a mix of activity groups (colors) in almost every session, which indicates that all major features of the interface have been leveraged broadly. As seed publications were already given in the study (mainly to save time by providing a starting point), activities to find and inspect them are limited and respective marks at the beginning of sessions do not show. Systematically inspecting the \suggested publications typically dominated the activities in the first halves of the sessions, while exploring the \network was rather salient in the second halves, especially when building a new literature collection~(\scNew). For updating a collection~(\scExisting), some participants \added{(e.g., PA1, PA2, PA3, PC1, PC2)} also started by exploring the network (usually in timeline mode), before then working through the suggested publications more systematically. \added{While others switched to the list-based representation for this, PA1 stayed within the network view.} For identifying experts (\scExpert), we also observe cases where participants \added{(e.g., PA2, PB1, PC1)} partly skip the extension of the collection and try to only work with the given collection. Activities related to introducing and adapting the \kwords are spread across the sessions, usually as shorter events. They are used by all participants for building a new collection (\scNew), but not as consistently for the other two scenarios (\deleted{\scExisting, \scExpert; two participants}\added{PB2, PB3 in \scExisting and PB2, PB4 in \scExpert} did not use keywords\deleted{ each}). Some of these observed strategies, we can link with answers from the questionnaire. For instance, systematically working through the top ranked suggested was mentioned by four participants \added{(PA3, PB2, PB3, PB4)}, potentially combined with filtering (\added{PA3, PB2}\deleted{two participants}), which aligns with larger periods of working with suggested publications. Moreover, timeline-based inspection was described by two participants \added{(PA1, PC1)} for updating a literature collection (\scExisting) and reflects in network phases at the beginning of the sessions.

\paragraph{Usage Patterns} On a finer temporal granularity, we also came across noteworthy repeated usage patterns. For dealing with \suggested publications, we analyzed the usage of filters and found examples in $20$ of $27$ sessions. Especially, a temporal filter was popular for updating an existing collection (\scExisting), with all participants but PA1using it. While, of course, everybody added publications to the selection to fulfill the tasks, it was not necessary to use the exclusion feature. However, except for PA1 and PC1, all participants at least used it once in a session, but rarely extensively. In the \network, participants applied different node settings either to experiment or to follow a specific goal. For instance, we observed at least five participants (PA2, PB1, PB2, PB3, PC2) who adjusted the number of nodes to get a clearer picture of potential experts (mostly as part of \scExpert). Finally, the \kwords feature helped in refining the ranking of publications and was repeatedly applied for building a new collection (\scNew; e.g., PA2, PB1, PB4). Multiple users (e.g., PA1, PA2, PB1, PB2, PC1)  dragged keyword nodes in the network, but usually were not satisfied with the result.

\paragraph*{Summary and Reflection} 
The results clearly demonstrate the practical applicability of \toolname to realistic literature search tasks. After a brief introduction, all participants---though having different levels of expertise---were able to work with the system successfully. They broadly confirmed the effective implementation of the targeted user stories. The observed search strategies covered all major interface features and aligned with the general layout of the tool---starting with some \selected publications, then extending the selection by working through \suggested publication and finally exploring the \network, while adapting and refining \kwords along the way.
Potential for improvement mainly showed for identifying experts (\scExpert), with lower ratings of the respective user story (\us{color_network}{4c}), lesser user priority compared to the other scenarios, and specific suggestions for improvement. While the timeline mode of the network received positive comments and ratings (\us{color_network}{4b}), the cluster mode was not perceived as positively (\us{color_network}{4a}) and could be still improved (e.g., to show clearer clusters). Regarding different levels of experience of users, as grouped in \Cref{fig:sessions}, we did not observe clear trends, but diverse individual usage behavior with specific patterns typically spread across all levels.

\subsection{Limitations}

A limitation of both the recommendation quality analysis and the user study was that the tested examples are confined to the visualization community.
Evaluating the quality of the recommendations could provide different results for different research areas. 
The effect of the \kwords depends on the chosen specific terms. Furthermore, the fit ratings in the study were subjectively judged by the author, however, transparently documented in the supplemental materials with additional comments clarifying negative decisions.
The user study, with its focus on user feedback and a detailed analysis of user behavior, lacks a quantitative perspective on tasks performance and general usability. 
Without involving a larger number of participants, all reported numeric user ratings are only considered preliminary, but were reported to provide context for the qualitative analysis.
As we recruited participants from our professional network, social desirability (to give favorable or specifically critical answers) might have influenced the absolute rankings of the user stories (see \Cref{tab:user_stories}), but does not affect their relative comparison. Being students and researchers related to visualization, the users do not well represent the full breadth of users from various disciplines.
Although participants had different experience with literature search and using \toolname, we did not study the long-term usage of the tool. \deleted{Neither did we test the effect of different levels of experience on the usage behavior.} 
A comparative analysis of different levels of academic experience is limited by the small number of users per level, and results are inconclusive.
Moreover, we cannot draw any conclusions whether \toolname performs better or worse than related approaches, beyond the variants of the different rankings tested and the feature comparison provided in the supplemental materials. Larger comparative studies are necessary, for instance, to contrast different citation-based recommendation approaches with each other, or to alternative approaches like traditional keyword search and recent alternatives leveraging generative artificial intelligence. \added{Such comparison should not only focus on task completion times and accuracy, but also consider to what extent users trust the recommendations, understood the results, and discovered relevant but potentially underrepresented works.}

\section{Discussion}
\label{sec:discussion}

Reflecting the results of the implementation and evaluation of \toolname, we identify the clear strengths of the approach, but also see certain weaknesses and ideas for improvement.

\paragraph{Applicability}
\toolname can be applied across research fields, as well as runs on various devices (responsiveness; see \Cref{sec:implementation}). Concerning tests in the area of visualization research, the recommendation quality evaluation and user study confirm that \toolname makes relevant recommendations. The novel keyword-based ranking has demonstrated to steer and improve recommendations. 
However, the recommendations can only be as complete as the underlying citation data. While we have not observed bigger gaps in the surveyed areas of visualization, other scientific communities might not be as well covered. Although the number of articles in \textit{Crossref} is growing rapidly, unfortunately, the percentage of publications with openly available references---the basis for citation analysis---has remained partial over the past years (around 60\% for journal articles)~\cite{10.31222/osf.io/smxe5}. However, campaigns like \textit{The Initiative for Open Citations} (I4OC) are supported by renown  institutions, which nurtures hope that data gaps will be closed. While the growing size of the underlying databases is appreciated and can be accommodated, the number of publications that the interface can handle is restricted. In \Cref{sec:eval:quality}, we have tested scenarios up to $84$ selected publications, and the interface stayed sufficiently performant on a regular laptop PC. However, on this scale, the citation network already tends to become cluttered and animations were not as smooth anymore. \added{To improve scalability, future enhancements could integrate approaches that split the network view into different perspectives to allow a more detailed exploration~\cite{10.3390/informatics4020011, 10.1109/tvcg.2006.166}. But such extensions might already go beyond the targeted scenarios (\scNew, \scExisting, \scExpert)---currently, our approach is not meant to survey or analyze larger corpora of literature.}
\deleted{While further incremental improvements could lead to somewhat better scalability, going beyond $200$ \selected publication will probably require different approaches, both visually and computationally, but also would be beyond the scope of the targeted scenarios (\scNew, \scExisting, \scExpert).}

\paragraph{Values}
The initially formulated values proved challenging to fulfill, but we claim to have made certain steps towards these goals. Regarding transparency of the recommendations (\vtransparancy), \toolname clarifies the rankings visually through glyphs. As users evaluated the related story (\us{color_suggested}{3a}) positively and made additional supportive comments on the recommendation approach, these efforts appear to be successful. The main innovation regarding user-centered control of the foraging process (\vcontrol) is the \kwords input to steer the ranking. While an effect shows, especially in earlier stages of literature exploration~(\scExisting), it seems less important when already a substantial amount of literature is collected (\scExisting). 
Additionally, users worked with the keywords frequently and rated the related user story positively (\us{color_keyword}{2}).
The most difficult to address, however, might be to counterbalance potential biases in the data and treat publications and authors fairly~(\vfairness). When trying to highlight potentially underrepresented literature, we can point out that they are yet \textit{unnoted}, but are not sure if this draws positive attention to the respective publications. The keywords did not rank up unnoted publications as initially expected. In contrast, we see substantially improved rankings regarding co-author diversity and early careers when favoring new and first-author publications. But our approach might still get affected through unfair practices, for instance, intentionally publishing more papers in smaller granularity for increased citations.
More empirical work, both data- and user-oriented, would be required to take next steps evaluating these values.

\section{Conclusion and Future Directions}
\label{sec:conclusion}

With \toolname, we contribute to a new generation of publication discovery interfaces. The unique focus is the transparent ranking of suggestions (\vtransparancy) and the control of suggestions through keywords (\vcontrol). The first is mainly achieved through a simple, understandable citation-based recommendation mechanism that is visually explained in consistently applied glyphs. The latter has successfully improved recommendations and contextualized the citation network. We also address biases in citation data (\vfairness) by highlighting potentially underrepresented works and promoting diversity in author rankings. However, these are small steps, and more research is needed for more inclusive recommendation approaches. Based on user feedback, the system appears ready for building new literature collections (\scNew) and updating or assessing existing one (\scExisting), but we still aim to invest more in features for relating collected publications to authors (\scExpert). Moreover, detailed empirical investigations are necessary to better understand the needs and preferences of the targeted diverse set of users, as well as to comparatively evaluate different visual literature exploration approaches. 

\acknowledgments{
The author wishes to thank \textbf{Solveig Rabsahl} for her substantial support in designing and conducting the user study, which also included the development of tailored evaluation visualizations; these contributions haven been funded by University of Bamberg through internal project funding. Furthermore, the author acknowledges contributions by \textbf{Cedric Krause} to an earlier version of \toolname, as documented in a previous poster presentation~\cite{10.2312/evp.20221110}. Thanks also to all early users who provided feedback and the participants of the user study.}

\section*{Supplemental Materials Index}
The supplemental materials are available at \url{https://osf.io/94ebr/} and include the following content:

\begin{itemize}
    \item A video demonstration of \toolname
    \item A feature-based comparison with related tools
    \item The bibliography as \BibTeX and \toolname session file
    \item Materials and results of the recommendation quality analysis (\Cref{sec:eval:quality})
    \item Materials and results of the user study (\Cref{sec:evalu:usage_study})
\end{itemize}

\bibliographystyle{abbrv-doi-hyperref}

\bibliography{references_all}
\end{document}